\documentclass[12pt, a4paper]{article}
\usepackage{geometry,amsmath,amssymb,amsthm,amsxtra,overpic,bbm,bm,epsfig,subfigure,ulem}
\usepackage{hyperref}
\usepackage{mathrsfs}
\usepackage{graphicx}
\usepackage{color}
\usepackage{comment}
\usepackage{epstopdf}
\usepackage{float}
\usepackage{cite}

\usepackage{slashed,stmaryrd}

\usepackage{multirow}

\usepackage{makecell}

\def\thefootnote{\fnsymbol{footnote}}

\newcommand{\eig}[2][q]{\lambda^{\rm #1}_{#2}}
\newcommand{\Aq}{A^{}_{\rm q}}

\newcommand{\muc}{\displaystyle \frac{m^{}_{u}}{m^{}_{c}}}
\newcommand{\mct}{\displaystyle \frac{m^{}_{c}}{m^{}_{t}}}
\newcommand{\mds}{\displaystyle \frac{m^{}_{d}}{m^{}_{s}}}
\newcommand{\msb}{\displaystyle \frac{m^{}_{s}}{m^{}_{b}}}

\newcommand{\msd}{\displaystyle \frac{m^{}_{s}}{m^{}_{d}}}

\newcommand{\eu}{\eta^{}_{\rm u}}
\newcommand{\ed}{\eta^{}_{\rm d}}
\newcommand{\ru}[1]{r^{#1}_{\rm u}}
\newcommand{\rd}[1]{r^{#1}_{\rm d}}

\begin{document}

\vspace{0.2cm}

\begin{center}
{\large\bf Correlations between quark mass and flavor mixing hierarchies}
\end{center}

\vspace{0.2cm}

\begin{center}
{\bf Harald Fritzsch$^{1}$},
{\bf Zhi-zhong Xing$^{2,3}$},
{\bf Di Zhang$^{2}$}
\footnote{E-mail: zhangdi@ihep.ac.cn (corresponding author)} \\
{\small $^{1}$Physics Department, Ludwig Maximilians University, D-80333 Munich,Germany \\
$^{2}$Institute of High Energy Physics and School of Physical Sciences, \\
University of Chinese Academy of Sciences, Beijing 100049, China \\
$^{3}$Center of High Energy Physics, Peking University, Beijing 100871, China}
\end{center}

\vspace{2cm}
\begin{abstract}
We calculate the quark flavor mixing matrix $V$ based
on the Hermitian quark mass matrices $M^{}_{\rm u}$ and $M^{}_{\rm d}$ with
vanishing $(1,1)$, $(1,3)$ and $(3,1)$ entries. The popular leading-order
prediction $|V^{}_{ub}/V^{}_{cb}| \simeq \sqrt{m^{}_u/m^{}_c}$ is significantly 
modified, and the result agrees with the current experimental value.
We find that behind the strong {\it mass} hierarchy
of up- or down-type quarks is the weak {\it texture} hierarchy of
$M^{}_{\rm u}$ or $M^{}_{\rm d}$ characterized by an approximate seesaw-like
relation among its $(2,2)$, $(2,3)$ and $(3,3)$ elements.
\end{abstract}

\def\thefootnote{\arabic{footnote}}
\setcounter{footnote}{0}

\newpage

\section{Introduction}

Since the standard model (SM) of particle physics was established in the 1960s
and 1970s, its flavor sector has never been fully understood in the sense that
all the flavor parameters are theoretically undetermined and their values
have to be extracted from various experimental measurements. Such a situation
will improve if the flavor structures of leptons and quarks can be well
constrained, e.g., with the help of certain flavor symmetries
\cite{Fritzsch:1999ee,Xing:2019vks}. In this connection the ``texture-zero"
approach, which was first developed in 1977 to calculate the Cabibbo angle of
quark flavor mixing \cite{Fritzsch:1977za,Weinberg:1977hb,Wilczek:1977uh},
is popular and helpful. No matter whether the vanishing entries in a fermion mass
matrix originate from a proper choice of the flavor basis or are enforced by
a kind of underlying flavor symmetry, they can in practice help establish some
testable relations between flavor mixing angles and ratios of fermion
masses. The key point is simply that flavor mixing is a measure of the intrinsic
mismatch between mass and flavor eigenstates of quarks, and thus
the Cabibbo-Kobayashi-Maskawa (CKM) quark mixing matrix $V$
\cite{Cabibbo:1963yz,Kobayashi:1973fv} may correlate with the respective mass
ratios of quarks via their texture zeros.

Today the values of six quark masses, extracted from quite a number of different
measurements~\cite{Zyla:2020zbs}, have been renormalized to the scale
$M^{}_Z = 91.1876 ~{\rm GeV}$ in the SM
framework~\cite{Xing:2007fb,Xing:2011aa,Huang:2020hdv}:
\begin{eqnarray}
&& m^{}_u = 1.23 \pm 0.21 ~{\rm MeV} \; , \quad
m^{}_c = 0.620 \pm 0.017 ~{\rm GeV} \; , \quad
m^{}_t = 168.26 \pm 0.75 ~{\rm GeV} \; ; \hspace{0.5cm}
\nonumber \\
&& m^{}_d = 2.67 \pm 0.19 ~{\rm MeV} \; , \quad
m^{}_s = 53.16 \pm 4.61 ~{\rm MeV} \; , \quad
m^{}_b = 2.839 \pm 0.026 ~{\rm GeV} \; .
\label{eq:data-mass}
\end{eqnarray}
On the other hand, a global fit of current experimental data on quark flavor mixing
and CP violation has provided us with the magnitudes of nine CKM matrix
elements to a very high degree of accuracy~\cite{Zyla:2020zbs}:
\begin{eqnarray}
&& \left|V^{}_{ud}\right| = 0.97401 \pm 0.00011 \; , \quad
\left|V^{}_{us}\right| = 0.22650 \pm 0.00048 \; , \quad
\left|V^{}_{ub}\right| = 0.00361^{+0.00011}_{-0.00009} \; , \hspace{0.5cm}
\nonumber \\
&& \left|V^{}_{cd}\right| = 0.22636 \pm 0.00048 \; , \quad
\left|V^{}_{cs}\right| = 0.97320 \pm 0.00011 \; , \quad
\left|V^{}_{cb}\right| = 0.04053^{+0.00083}_{-0.00061} \; , \hspace{0.5cm}
\nonumber \\
&& \left|V^{}_{td}\right| = 0.00854^{+0.00023}_{-0.00016} \; , \quad
\left|V^{}_{ts}\right| = 0.03978^{+0.00082}_{-0.00060} \; , \quad
\left|V^{}_{tb}\right| = 0.999172^{+0.000024}_{-0.000035} \; .
\label{eq:data-mixing}
\end{eqnarray}
How to link the hierarchical pattern of $V$ to the hierarchical mass spectra of
up- and down-type quarks is therefore a burning question.

In this regard one may make a survey of all the
phenomenologically allowed zero textures of Hermitian quark mass matrices (see,
e.g., Refs.~\cite{Ramond:1993kv,Peccei:1995fg,Zhou:2003ji,Bagai:2021nsl} for very
comprehensive studies)
\footnote{Without loss of generality, the up- and down-type quark mass matrices
can always be taken to be Hermitian after a proper choice of the flavor basis in
the SM or its extensions which have no flavor-changing right-handed currents
\cite{Frampton:1985qk}. A limited number of texture zeros can actually be achieved
in this way \cite{Branco:1988iq}.}.
Then it is easy to confirm that the texture \cite{Du:1992iy,Fritzsch:1995nx,Xing:1996hi}
\begin{eqnarray}
M^{}_{\rm q} =
\begin{pmatrix}
0 & C^{}_{\rm q} & 0
\\
C^\ast_{\rm q} & \tilde{B}^{}_{\rm q} & B^{}_{\rm q}
\\
0 & B^\ast_{\rm q} & A^{}_{\rm q}
\end{pmatrix} \;
\label{eq:mass-matrix}
\end{eqnarray}
is of particular interest, where $\rm q = u$ (up) or d (down) and
$A^{}_{\rm q} > 0$ can always be arranged.
Note that this texture is actually the most natural extension of the original
Fritzsch texture \cite{Fritzsch:1977vd,Fritzsch:1979zq,Fritzsch:1986sn}
by allowing for $\tilde{B}^{}_{\rm q} \neq 0$.
Since the structures of $M^{}_{\rm u}$
and $M^{}_{\rm d}$ in Eq.~(\ref{eq:mass-matrix}) are exactly parallel, they
should originate from the same underlying flavor dynamics. After making the unitary
transformation $V^\dagger_{\rm q} M^{}_{\rm q} V^{}_{\rm q} = {\rm diag}
\{\lambda^{\rm q}_1 , \lambda^{\rm q}_2 , \lambda^{\rm q}_3\}$ with
$\lambda^{\rm q}_i$ being the quark mass eigenvalues (for $i = 1, 2, 3$),
one may obtain the CKM matrix $V \equiv V^\dagger_{\rm u} V^{}_{\rm d}$.
The structural parallelism between $M^{}_{\rm u}$ and $M^{}_{\rm d}$ implies
a structural parallelism between $V^{}_{\rm u}$ and $V^{}_{\rm d}$, and thus
$V$ is expected to be close to the identity matrix $I$. This expectation is
certainly consistent with the observed pattern of $V$, which deviates from
$I$ only at the level of ${\cal O}(20\%)$ \cite{Zyla:2020zbs}. Another
remarkable merit of the zero texture in Eq.~(\ref{eq:mass-matrix}) is its
analytical calculability, which helps us to exactly express the elements of
$V^{}_{\rm q}$ in terms of $\lambda^{\rm q}_i$ and $A^{}_{\rm q}$
\cite{Xing:2003yj} on the one hand and make some reasonable analytical
approximations on the other hand.

Although quite a lot of attention has been paid to the above textures of
$M^{}_{\rm u,d}$ and their phenomenological consequences, we realize
that in this connection reliable analytical approximations up to a sufficiently
good degree of accuracy are still lacking. In particular, the {\it failure}
of the popular leading-order prediction
$|V^{}_{ub}/V^{}_{cb}| \simeq \sqrt{m^{}_u/m^{}_c}$ against current
experimental data has not been analytically resolved. Here we are going to
fill in this gap and understand why behind the strong mass hierarchy
of up- or down-type quarks is the weak texture hierarchy of $M^{}_{\rm u}$
or $M^{}_{\rm d}$ characterized by an approximate seesaw-like relation
$\tilde{B}^{}_{\rm q} \sim |B^{}_{\rm q}|^2/A^{}_{\rm q}$
(for $\rm q = u$ or $\rm d$).
Such insights will be very helpful for a much deeper study of the quark flavor issues.

\section{Analytical approximations}
\label{section:2}

One may transform $M^{}_{\rm q}$ in Eq.~(\ref{eq:mass-matrix}) into a real
symmetric matrix $\overline{M}^{}_{\rm q}$ by a phase redefinition:
\begin{eqnarray}
\overline{M}^{}_{\rm q} = P^{}_{\rm q} M^{}_{\rm q} P^\dagger_{\rm q} =
\begin{pmatrix}
0 & | C^{}_{\rm q} | & 0
\\
| C^{}_{\rm q} | & \tilde{B}^{}_{\rm q} & |B^{}_{\rm q}|
\\
0 & |B^{}_{\rm q}| & A^{}_{\rm q}
\end{pmatrix} \;,
\label{eq:mass-matrix-real}
\end{eqnarray}
where $P^{}_{\rm q} = {\rm diag} \{ 1, e^{{\rm i} \phi^{}_{\rm q}},
e^{{\rm i} \phi^\prime_{\rm q}} \}$ with $\phi^{}_{\rm q} \equiv {\rm arg}
\left( C^{}_{\rm q} \right)$ and $\phi^\prime_{\rm q} \equiv {\rm arg}
\left( B^{}_{\rm q} \right) + {\rm arg} \left( C^{}_{\rm q} \right)$.
It is obvious that $\overline{M}^{}_{\rm q}$ can be diagonalized
via an orthogonal transformation $O^T_{\rm q} \overline{M}^{}_{\rm q}
O^{}_{\rm q} = {\rm diag} \{ \lambda^{\rm q}_1, \lambda^{\rm q}_2,
\lambda^{\rm q}_3 \}$ with $\lambda^{\rm q}_i$ denoting the three eigenvalues
of $M^{}_{\rm q}$. Then $\tilde{B}^{}_{\rm q}$, $|B^{}_{\rm q}|$ and
$|C^{}_{\rm q}|$ can be expressed in terms of $A^{}_{\rm q}$ and
$\lambda^{\rm q}_i$ as follows~\cite{Fritzsch:2002ga}:
\begin{eqnarray}
\tilde{B}^{}_{\rm q} &=& \lambda^{\rm q}_1 + \lambda^{\rm q}_2 +
\lambda^{\rm q}_3 - A^{}_{\rm q} \;,
\nonumber
\\
|B^{}_{\rm q}|^2 &=& \frac{\left( A^{}_{\rm q} - \lambda^{\rm q}_1 \right)
\left( A^{}_{\rm q} - \lambda^{\rm q}_2 \right) \left( \lambda^{\rm q}_3 -
A^{}_{\rm q} \right) }{A^{}_{\rm q}} \;,
\nonumber
\\
|C^{}_{\rm q}|^2 &=& \frac{-\lambda^{\rm q}_1 \lambda^{\rm q}_2
\lambda^{\rm q}_3}{A^{}_{\rm q}} \;.
\label{eq:mass-para}
\end{eqnarray}
Note that the strong hierarchy
$|\lambda^{\rm q}_1| \ll |\lambda^{\rm q}_2| \ll |\lambda^{\rm q}_3|$
allows us to choose $\lambda^{\rm q}_3 >0$ in correspondence to $A^{}_{\rm q}>0$,
and thus $\lambda^{\rm q}_1 \lambda^{\rm q}_2 < 0$ as required by
$\det\overline{M}^{}_{\rm q} = -A^{}_{\rm q} |C^{}_{\rm q}|^2 =
\lambda^{\rm q}_1 \lambda^{\rm q}_2 \lambda^{\rm q}_3$. In this convention
the orthogonal matrix $O^{}_{\rm q}$ is exactly given by
\small
\begin{eqnarray}
O^{}_{\rm q} =
\begin{pmatrix}
\displaystyle \sqrt{\frac{\eig{2} \eig{3} \left( \Aq - \eig{1} \right)  }{\Aq
\left( \eig{2} - \eig{1} \right) \left( \eig{3} - \eig{1} \right) }} &
\displaystyle \eta^{}_{\rm q} \sqrt{\frac{ \eig{1} \eig{3} \left( \eig{2} -
\Aq \right) }{\Aq \left( \eig{2} - \eig{1} \right) \left( \eig{3} - \eig{2}
\right)}} & \displaystyle \sqrt{\frac{ \eig{1} \eig{2} \left( \Aq - \eig{3}
\right)}{\Aq \left( \eig{3} - \eig{1} \right) \left( \eig{3} - \eig{2} \right) }}
\vspace{0.2cm}\\
\displaystyle - \eta^{}_{\rm q} \sqrt{\frac{\eig{1}\left( \eig{1} - \Aq \right)}
{\left( \eig{2} - \eig{1} \right) \left( \eig{3} - \eig{1} \right)}} & \displaystyle
\sqrt{\frac{\eig{2}\left( \Aq - \eig{2} \right)}{\left( \eig{2} - \eig{1} \right)
\left( \eig{3} - \eig{2} \right)}} & \displaystyle \sqrt{\frac{\eig{3} \left(
\eig{3} - \Aq \right)}{\left( \eig{3} - \eig{1} \right) \left( \eig{3} - \eig{2}
\right)}}
\vspace{0.2cm}\\
\displaystyle \eta^{}_{\rm q} \sqrt{\frac{\eig{1} \left( \Aq - \eig{2} \right)
\left( \Aq - \eig{3} \right)}{\Aq \left( \eig{2} - \eig{1} \right) \left( \eig{3} -
\eig{1} \right)}} & \displaystyle ~ - \sqrt{\frac{\eig{2}\left(\Aq - \eig{1}\right)
\left( \eig{3} - \Aq \right)}{\Aq \left( \eig{2} - \eig{1} \right) \left( \eig{3} -
\eig{2} \right)}} ~ & \displaystyle \sqrt{\frac{\eig{3} \left(\Aq - \eig{1} \right)
\left( \Aq - \eig{2} \right)}{\Aq \left( \eig{3} - \eig{1} \right) \left( \eig{3} -
\eig{2} \right)}}
\end{pmatrix} \;\;\;
\label{eq:orth}
\end{eqnarray}
\normalsize
with $\eta^{}_{\rm q} = + 1$ and $\eta^{}_{\rm q} = - 1$ corresponding
respectively to $\eig{2}>0$ and $\eig{2} < 0$. So the explicit relations between
$\eig{i}$ and real quark masses are
\footnote{One may consider eliminating the sign uncertainty associated with
$\eta^{}_{\rm q}$ and thus arrange all the three mass eigenvalues of
$\overline{M}^{}_{\rm q}$ (for $\rm q = u$ or $\rm d$) to be positive by
properly redefining the phases of three right-handed quark fields. Such a
treatment is equivalent to a new choice of the flavor basis for $M^{}_{\rm q}$,
and that is why the orthogonal matrix $O^{}_{\rm q}$ used to diagonalize
$\overline{M}^{}_{\rm q}$ will be accordingly affected.
So different choices of the values of $\eta^{}_{\rm u}$ and
$\eta^{}_{\rm d}$ mean our considerations of somewhat different quark mass
matrices (i.e., their corresponding nonzero elements are not exactly equal
to one another) --- this point will become transparent in our numerical results.}
\begin{eqnarray}
\big(\eig[u]{1}, \eig[u]{2}, \eig[u]{3} \big) &=&
\big(-\eu m^{}_u, \eu m^{}_c, m^{}_t \big) \; ,
\nonumber \\
\big(\eig[d]{1}, \eig[d]{2}, \eig[d]{3} \big) &=&
\big(-\ed m^{}_d, \ed m^{}_s, m^{}_b \big) \; , \hspace{0.5cm}
\label{eq:mass}
\end{eqnarray}
for the up- and down-quark sectors, respectively. As a consequence, nine
elements of the CKM matrix $V = O^T_{\rm u}
P^{}_{\rm u} P^\dagger_{\rm d} O^{}_{\rm d}$ can be expressed as
\begin{eqnarray}
V^{}_{pq} = \left( O^{}_{\rm u} \right)^{}_{1p} \left( O^{}_{\rm d} \right)^{}_{1q}
+ \left( O^{}_{\rm u} \right)^{}_{2p} \left( O^{}_{\rm d} \right)^{}_{2q}
e^{{\rm i} \phi^{}_1} + \left( O^{}_{\rm u} \right)^{}_{3p} \left( O^{}_{\rm d}
\right)^{}_{3q} e^{{\rm i} \phi^{}_2} \;,
\label{eq:Vckm}
\end{eqnarray}
where $p$ and $q$ run respectively over the flavor indices $(u, c, t)$ and $(d, s, b)$,
and the two phase differences are defined as
\begin{eqnarray}
\phi^{}_1 &\equiv& \phi^{}_{\rm u} - \phi^{}_{\rm d}
= \arg\left(C^{}_{\rm u}\right) - \arg\left(C^{}_{\rm d}\right) \; ,
\nonumber \\
\phi^{}_2 &\equiv& \phi^{\prime}_{\rm u} - \phi^{\prime}_{\rm d}
= \arg\left(B^{}_{\rm u}\right) - \arg\left(B^{}_{\rm d}\right) + \phi^{}_1 \; .
\hspace{0.5cm}
\label{eq:phase}
\end{eqnarray}
If the values of six quark masses are input, one may determine the elements of $V$ and
check their consistency with current experimental data on flavor mixing and CP violation
by adjusting the four free parameters $A^{}_{\rm u}$, $A^{}_{\rm d}$, $\phi^{}_1$ and
$\phi^{}_2$. In this regard a careful numerical analysis of the allowed parameter space
has been done in Ref.~\cite{Xing:2015sva}. Here we are going to carry out a careful
analytical exploration of the salient features of $V$ by making reliable approximations
for $O^{}_{\rm u}$ and $O^{}_{\rm d}$ in Eq.~(\ref{eq:orth}) {\it for the first time},
so as to truly understand the link between the texture zeros of $M^{}_{\rm q}$ and the
observed pattern of $V$ in depth.

To assure that our analytical approximations are reliable, we need to take
into account the quark masses and CKM matrix elements in a common
energy scale as given in Eqs.~(\ref{eq:data-mass}) and (\ref{eq:data-mixing}).
One can easily see $m^{}_u/m^{}_c \sim m^{}_c/m^{}_t \sim \lambda^4$ and $m^{}_d/m^{}_s
\sim m^{}_s/m^{}_b \sim \lambda^2$, together with $|V^{}_{cd}| \simeq |V^{}_{us}|
\equiv \lambda$, $|V^{}_{ts}| \simeq |V^{}_{cb}| \sim \lambda^2$, $|V^{}_{td}|
\sim \lambda^3$ and $|V^{}_{ub}| \sim |V^{}_{td}|/2$, where $\lambda$ is defined as
a small expansion parameter for the CKM matrix $V$. The much stronger hierarchy of the
up-type quark masses implies that their ratios contribute much less to the elements
of $V$. In other words, the CKM flavor mixing parameters are expected to be dominated
by those contributions from the down-quark sector.

\subsection{Generic approximations}

To measure how hierarchical the four entries on the right-bottom corner of
$M^{}_{\rm q}$ can be in fitting the present experimental data, let us define
the following two characteristic parameters:
\begin{eqnarray}
\ru{} \equiv 1 - \frac{A^{}_{\rm u}}{m^{}_t} \; , \quad
\rd{} \equiv 1 - \frac{A^{}_{\rm d}}{m^{}_b} \; .
\label{eq:r}
\end{eqnarray}
The careful numerical analysis made in Ref.~\cite{Xing:2015sva} indicates that
the allowed parameter space of $\ru{}$ and $\rd{}$ is mainly located in the range
of $0.1$ to $0.2$, although there is also a small possibility of
$\ru{} \sim \rd{} \sim 0.5$. In this case we typically assume
$r^{}_{\rm u} \sim r^{}_{\rm d}\sim \mathcal{O} \left( \lambda \right)$ when
making our analytical approximations
\footnote{The accuracy of our analytical approximations is found to be only
slightly worse in the case of $\ru{} \sim \rd{} \sim 0.5$.}.
Up to the accuracy of $\mathcal{O}\left( \lambda^4 \right)$, we have
\allowdisplaybreaks[4]
\begin{eqnarray}
\left(O^{}_{\rm u}\right)^{}_{1 u} &\simeq&
1 -  \displaystyle\frac{1}{2} \muc  \; ,
\nonumber \\
\left(O^{}_{\rm u}\right)^{}_{1 c} &\simeq& \eu \sqrt{\muc} \; ,
\nonumber \\
\left(O^{}_{\rm u}\right)^{}_{1 t} &\simeq& 0 \; ,
\nonumber \\
\left(O^{}_{\rm u}\right)^{}_{2 u} &\simeq&
\displaystyle -\eu\sqrt{\muc} \sqrt{1 - \ru{}} \; ,
\nonumber \\
\left(O^{}_{\rm u}\right)^{}_{2 c} &\simeq&
\displaystyle \sqrt{1 - \ru{} } - \frac{1}{2} \muc \; ,
\nonumber \\
\left(O^{}_{\rm u}\right)^{}_{2 t} &\simeq&
\displaystyle \sqrt{\ru{}} \; ,
\nonumber \\
\left(O^{}_{\rm u}\right)^{}_{3 u} &\simeq&
\displaystyle\eu \sqrt{\ru{}\muc} \; ,
\nonumber \\
\left(O^{}_{\rm u}\right)^{}_{3 c} &\simeq&
\displaystyle -\sqrt{\ru{}} \; ,
\nonumber \\
\left(O^{}_{\rm u}\right)^{}_{3 t} &\simeq&
\displaystyle \sqrt{1 - \ru{}} \; ;
\label{eq:app-u}
\end{eqnarray}
and
\begin{eqnarray}
\left(O^{}_{\rm d}\right)^{}_{1 d} &\simeq&
\displaystyle 1 - \frac{1}{2} \mds \left( 1 - \frac{3}{4} \mds\right) \; ,
\nonumber \\
\left(O^{}_{\rm d}\right)^{}_{1 s} &\simeq&
\displaystyle \ed \sqrt{\mds} \left( 1 -\frac{1}{2}
\mds - \frac{1}{2} \ed \rd{} \msb \right) \; ,
\nonumber \\
\left(O^{}_{\rm d}\right)^{}_{1 b} &\simeq&
\displaystyle \sqrt{\rd{}} \msb \sqrt{\mds} \; ,
\nonumber \\
\left(O^{}_{\rm d}\right)^{}_{2 d} &\simeq&
\displaystyle -\ed \sqrt{\mds} \left[ \sqrt{1 - \rd{}}
- \frac{1}{2} \mds \left( 1 - \frac{1}{2} \rd{} \right) \right] \; ,
\nonumber \\
\left(O^{}_{\rm d}\right)^{}_{2 s} &\simeq&
\displaystyle \sqrt{1 - \rd{}} - \frac{1}{2} \mds \left( 1 - \frac{3}{4}
\mds \right) + \frac{1}{4} \rd{} \mds \left( 1 + \frac{1}{4} \rd{}
\right) - \frac{1}{2} \ed \rd{} \msb \left( 1 + \frac{1}{2} \rd{} \right) \; ,
\nonumber \\
\left(O^{}_{\rm d}\right)^{}_{2 b} &\simeq&
\displaystyle \sqrt{\rd{}} \left( 1 + \frac{1}{2} \ed \msb \right) \; ,
\nonumber \\
\left(O^{}_{\rm d}\right)^{}_{3 d} &\simeq&
\displaystyle \ed \sqrt{\rd{} \mds} \left[ 1 - \frac{1}{2}
\left( \mds + \ed \msb \right) \right] \; ,
\nonumber \\
\left(O^{}_{\rm d}\right)^{}_{3 s} &\simeq&
\displaystyle - \sqrt{\rd{}} \left[ 1 + \frac{1}{2} \left(\ed \msb - \mds \right)
\right] \; ,
\nonumber \\
\left(O^{}_{\rm d}\right)^{}_{3 b} &\simeq&
\displaystyle \sqrt{1-\rd{}} - \frac{1}{2} \ed \rd{} \msb
\left( 1 + \frac{1}{2} \rd{} \right) \; ,
\label{eq:app-d}
\end{eqnarray}
where $\sqrt{1- r^{}_{\rm q}}$ is only
written for short and it should be expanded as $\sqrt{1- r^{}_{\rm q}} \simeq 1 -
r^{}_{\rm q}/2 - r^2_{\rm q}/8 + \cdots$ in
order to assure that the relevant terms can reach the precision of
$\mathcal{O} \left( \lambda^4 \right)$.
Since the three up-type quarks have a much stronger
mass hierarchy, the analytical approximations made for the elements of $O^{}_{\rm u}$
are much simpler at the $\mathcal{O} \left( \lambda^4 \right)$ level as compared
with the corresponding results of $O^{}_{\rm d}$ to the same degree of accuracy.

With the help of Eqs.~(\ref{eq:Vckm}), (\ref{eq:app-u}) and (\ref{eq:app-d}),
we obtain the following approximate results for nine elements of the CKM matrix
$V$:
\begin{eqnarray}
V^{}_{ud} &\simeq& 1 - \frac{1}{2} \mds + \eu\ed e^{{\rm i} \phi^{}_1}
\sqrt{\muc} \sqrt{\mds} \;,
\nonumber
\\
V^{}_{us} &\simeq& \ed \sqrt{\mds} \left( 1 - \frac{1}{2} \mds \right)
+ \frac{1}{2} \eu e^{{\rm i}\phi^{}_1}
\sqrt{\muc} \left( \ru{} - 2\sqrt{\ru{}\rd{}} e^{-{\rm i} \delta} + \rd{} - 2 \right) \;,
\nonumber
\\
V^{}_{ub} &\simeq& \frac{1}{2} \eu{} \sqrt{\muc} \Big[ \left( 2 - \rd{} \right)
\sqrt{\ru{}} e^{{\rm i} \phi^{}_2} - \left( 2 - \ru{} \right) \sqrt{\rd{}} e^{{\rm i}
\phi^{}_1} \Big] + \sqrt{\rd{}} \msb \sqrt{\mds} \;,
\nonumber
\\
V^{}_{cd} &\simeq& \frac{1}{2} \ed e^{{\rm i}\phi^{}_1} \sqrt{\mds} \left[ \mds +
\frac{1}{4} \left( \ru{} - \rd{} \right)^2 + \ru{} - 2\sqrt{\ru{}\rd{}}
e^{-{\rm i}\delta} + \rd{} - 2 \right] + \eu \sqrt{\muc} \;,
\nonumber
\\
V^{}_{cs} &\simeq& e^{{\rm i} \phi^{}_1} \left[ 1 - \frac{1}{4} \left( 2 - \mds \right)
\left( \ru{} - 2\sqrt{\ru{}\rd{}} e^{-{\rm i}\delta} + \rd{} \right) - \frac{1}{16}
\left(\ru{} + \rd{} + 2 \right) \left( \ru{} - \rd{} \right)^2 \right.
\hspace{0.5cm}
\nonumber
\\
&& - \left. \frac{1}{2} \mds - \frac{1}{2} \ed \msb \left( \rd{} - \sqrt{\ru{}\rd{}}
e^{-{\rm i}\delta} \right) + \eu\ed e^{-{\rm i}\phi^{}_1} \sqrt{\muc} \sqrt{\mds}
\right] \;,
\nonumber
\\
V^{}_{cb} &\simeq& e^{{\rm i} \phi^{}_1} \left\{ \sqrt{\rd{}}
\left[ 1 - \frac{1}{2} \ru{} - \frac{1}{8} \ru{2} - \frac{1}{16} \ru{3}
+ \frac{1}{4} \ed \left( 2 - \ru{} \right) \msb \right] \right.
\nonumber
\\
&& - \left. \sqrt{\ru{}} e^{-{\rm i} \delta} \left( 1 - \frac{1}{2} \rd{}
- \frac{1}{8} \rd{2} - \frac{1}{16} \rd{3} - \frac{1}{2} \ed \rd{} \msb \right)
\right\} \;,
\nonumber
\\
V^{}_{td} &\simeq& \ed e^{{\rm i}\phi^{}_1} \sqrt{\mds} \left[ \sqrt{\rd{}}
e^{-{\rm i} \delta} \left( 1 - \frac{1}{2} \ru{} - \frac{1}{8} \ru{2} -
\frac{1}{2} \mds -\frac{1}{2} \ed \msb\right) \right.
\nonumber
\\
&& - \left. \sqrt{\ru{}} \left( 1 - \frac{1}{2} \rd{} - \frac{1}{8} \rd{2} -
\frac{1}{2} \mds \right) \right] \;,
\nonumber
\\
V^{}_{ts} &\simeq& e^{{\rm i} \phi^{}_1} \left\{ \sqrt{\ru{}}
\left[ 1 - \frac{1}{2} \rd{} - \frac{1}{8} \rd{2} - \frac{1}{16} \rd{3} -
\frac{1}{4} \left( 2 -\rd{} \right)\mds - \frac{1}{2} \ed \rd{} \msb \right] \right.
\nonumber
\\
&& - \left. \sqrt{\rd{}} e^{-{\rm i} \delta}
\left[ 1 - \frac{1}{2} \ru{} - \frac{1}{8} \ru{2} - \frac{1}{16} \ru{3} +
\frac{1}{4} \left( 2 - \ru{} \right) \left( \ed \msb - \mds \right) \right]
\right\} \;,
\nonumber
\\
V^{}_{tb} &\simeq& e^{{\rm i} \phi^{}_2} \left[ 1 - \frac{1}{2} \left( \ru{}
- 2\sqrt{\ru{}\rd{}} e^{{\rm i} \delta} + \rd{} \right) -
\frac{1}{16} \left( \ru{} + \rd{} + 2 \right) \left(\ru{} - \rd{} \right)^2 \right.
\nonumber
\\
&& - \left. \frac{1}{2} \ed \msb \left( \rd{} - \sqrt{\ru{}\rd{}} e^{{\rm i}
\delta} \right) \right] \; ,
\label{eq:ele-Vckm}
\end{eqnarray}
where $\delta \equiv \phi^{}_1 - \phi^{}_2 = \arg\left(B^{}_{\rm d}\right)
- \arg\left(B^{}_{\rm u}\right)$ is defined. Then the moduli of nine
CKM matrix elements $\left| V^{}_{pq} \right| = \sqrt{ V^{}_{pq} V^\ast_{pq}}$
(for $p = u, c, t$ and $q = d, s , b$) are found to be
\begin{eqnarray}
\left| V^{}_{ud} \right| &\simeq& 1 - \frac{1}{2} \mds + \eu\ed
\sqrt{\muc}\sqrt{\mds} \cos\phi^{}_1 \;,
\nonumber
\\
\left| V^{}_{us} \right| &\simeq& \left( 1 - \frac{1}{2} \mds \right) \sqrt{\mds} + \frac{1}{2}
\muc \sqrt{\msd} \sin^2\phi^{}_1 + \frac{1}{2} \eu\ed \sqrt{\muc} \Big[ \left( \ru{} + \rd{} -
2 \right) \cos\phi^{}_1
\nonumber
\\
&& - 2\sqrt{\ru{}\rd{}} \cos\phi^{}_2 \Big] \;,
\nonumber
\\
\left| V^{}_{ub} \right| &\simeq& \frac{1}{2\sqrt{R}} \sqrt{\muc} \Big[ 2R - 2\ru{}\rd{} +
\left( \ru{} + \rd{} \right) \sqrt{\ru{}\rd{}} \cos\delta \Big] - \frac{1}{\sqrt{R}} \eu
\msb \sqrt{\mds}
\nonumber
\\
&& \times \left( \rd{} \cos\phi^{}_1 - \sqrt{\ru{}\rd{}} \cos\phi^{}_2 \right)  \;,
\nonumber
\\
\left| V^{}_{cd} \right| &\simeq& \sqrt{\mds} \left[ 1 - \frac{1}{2} R - \frac{1}{2} \mds +
\frac{1}{2} \ru{} \rd{} \sin^2 \delta - \frac{1}{8} \left( \ru{} - \rd{} \right)^2 \right] +
\frac{1}{2} \muc \sqrt{\msd} \sin^2\phi^{}_1
\nonumber
\\
&& - \eu \ed \sqrt{\muc} \left( \cos\phi^{}_1 + \sqrt{\ru{}\rd{}} \sin\phi^{}_1 \sin\delta
\right)\;,
\nonumber
\\
\left| V^{}_{cs} \right| &\simeq& 1 - \frac{1}{2} R - \frac{1}{4} \left( 2 - R \right) \mds
-\frac{1}{16} \Big\{ \ru{3} + \rd{3} + 2\left( \ru{} -\rd{} \right)^2 - \ru{}\rd{} \Big[
\left(3 - 2\cos2\delta \right)
\nonumber
\\
&& \times \left( \ru{} + \rd{} \right) + 4 \sin\delta \left( 2\sin\delta -
\sqrt{\ru{}\rd{}} \sin2\delta \right) \Big] \Big\} + \eu\ed \sqrt{\muc} \sqrt{\mds}
\cos{\phi^{}_1}
\nonumber
\\
&& - \frac{1}{2} \ed \msb \left( \rd{} - \sqrt{\ru{}\rd{}} \cos\delta \right) \;,
\nonumber
\\
\left| V^{}_{cb} \right| &\simeq& \sqrt{R} + \frac{1}{2\sqrt{R}}
\sqrt{\ru{}\rd{}} \Big[ \left( \ru{} + \rd{} \right) \cos\delta - 2\sqrt{\ru{}\rd{}}
\Big] + \frac{1}{2R\sqrt{R}} \Big\{\frac{1}{4}
\left(\cos2\delta - 3 \right) \ru{2}\rd{2}
\nonumber
\\
&& + \frac{1}{4} \left( \ru{} + \rd{} \right)^3 \sqrt{\ru{}\rd{}} \cos\delta - \frac{3}{4}
\ru{}\rd{} \left( \ru{2} + \rd{2} \right) \cos^2\delta + \ed \msb \Big[ \rd{2} - \left(
\ru{} + 3\rd{} \right)
\nonumber
\\
&& \times \sqrt{\ru{}\rd{}} \cos\delta + \ru{}\rd{} \left( 2 + \cos2\delta
\right) \Big] \Big\} \;,
\nonumber
\\
\left| V^{}_{td} \right| &\simeq& \frac{1}{2\sqrt{R}} \sqrt{\mds} \left\{ \left( 2 - \mds
\right) R + \sqrt{\ru{}\rd{}} \Big[ \left( \ru{} + \rd{} \right) \cos\delta -
2\sqrt{\ru{}\rd{}} \Big]  \right.
\nonumber
\\
&& - \frac{1}{4R} \ru{}\rd{} \Big[ \left( \ru{2} + \rd{2} \right) \cos^2\delta
+ \ru{}\rd{} \left( 5 + \cos2\delta \right) - 4 \sqrt{\ru{}\rd{}} \left( \ru{} + \rd{}
\right) \cos\delta \Big]
\nonumber
\\
&& + \left. \frac{1}{4} \left( \ru{} - \rd{} \right)^2 \sqrt{\ru{}\rd{}}
\cos\delta  - \ed \msb \left( \rd{} - \sqrt{\ru{}\rd{}} \cos\delta \right) \right\} \;,
\nonumber
\\
\left| V^{}_{ts} \right| &\simeq& \frac{1}{2} \sqrt{R} \left( 2 -\mds \right) +
\frac{1}{2\sqrt{R}} \sqrt{\ru{}\rd{}} \Big[ \left( \ru{} + \rd{} \right) \cos\delta -
2\sqrt{\ru{}\rd{}} \Big] + \frac{1}{8R\sqrt{R}}
\nonumber
\\
&& \times \Big\{ \sqrt{\ru{}\rd{}} \left( \ru{} + \rd{} \right)^3 \cos\delta - \ru{}\rd{}
\Big[ 3\left( \ru{} - \rd{} \right)^2 \cos^2\delta + 4\ru{}\rd{} \left( 1+\cos^2\delta
\right) \Big]
\nonumber
\\
&& + 4 \ed \msb \Big[ \rd{2} + \ru{}\rd{} \left( 2+\cos2\delta \right) -
\sqrt{\ru{}\rd{}} \left( \ru{} + 3\rd{} \right) \cos\delta \Big] \Big\} \;,
\nonumber
\\
\left| V^{}_{tb} \right| &\simeq& 1 - \frac{1}{2} R - \frac{1}{16} \left( \ru{} + \rd{} +
2 \right) \left( \ru{} - \rd{} \right)^2 + \frac{1}{4} \left( 2 + R \right) \ru{} \rd{}
\sin^2\delta - \frac{1}{2} \ed \msb
\nonumber
\\
&& \times \left( \rd{} - \sqrt{\ru{}\rd{}} \cos\delta \right) \;,
\label{eq:abs-Vckm}
\end{eqnarray}
where $R \equiv \ru{} + \rd{} - 2\sqrt{\ru{}\rd{}} \cos\delta$ has been defined.
Note that in Eq.~(\ref{eq:abs-Vckm}) the expansion terms of
$|V^{}_{cb}|$ and $|V^{}_{ts}|$ are only kept up to $\mathcal{O} \left( \lambda^{2.5}
\right)$ for the sake of simplicity, and the accuracy of the other CKM
matrix elements is at the level of $\mathcal{O} \left( \lambda^{3.5}\right)$.
Note also that $R \lesssim {\cal O}\left(\lambda\right)$ is required to assure
the analytical approximations made above to be valid. If the
experimental values of $|V^{}_{ub}|$, $|V^{}_{cb}|$, $|V^{}_{td}|$ and $|V^{}_{ts}|$
are taken into consideration, we actually find that $R$ should be
of $\mathcal{O} \left(\lambda^4 \right)$. One will see later on
that the smallness of $R$ mainly results from $\ru{} \simeq \rd{}$ and $\delta \sim 0$.

With the help of Eq.~(\ref{eq:abs-Vckm}), we immediately arrive at the following
three typical ratios of the CKM matrix elements:
\begin{eqnarray}
\left| \frac{V^{}_{ub}}{V^{}_{cb}} \right| &\simeq& \sqrt{\muc} + \eu \msb
\sqrt{\rd{}\mds} \frac{\sqrt{\ru{}} \cos\phi^{}_2 - \sqrt{\rd{}} \cos\phi^{}_1}{\ru{}
+ \rd{} - 2\sqrt{\ru{}\rd{}} \cos\delta} \;,
\nonumber
\\
\left| \frac{V^{}_{td}}{V^{}_{ts}} \right| &\simeq& \sqrt{\mds} \left[ 1 - \ed \msb
\frac{\rd{2} +\left( 2 + \cos2\delta \right)\ru{}\rd{} - \left( 3\rd{} + \ru{} \right)
\sqrt{\ru{}\rd{}} \cos\delta}{\left( \ru{} + \rd{} - 2\sqrt{\ru{}\rd{}} \cos\delta
\right)^2 } \right] \;,
\label{eq:Ratio}
\end{eqnarray}
and
\begin{eqnarray}
\left| \frac{V^{}_{us}}{V^{}_{ud}} \right| &\simeq& \sqrt{\mds} + \frac{1}{2} \eu\ed
\sqrt{\muc} \Big[ \left( \ru{} + \rd{} - 2 \right) \cos\phi^{}_1 -  2\sqrt{\ru{}\rd{}}
\cos\phi^{}_2 \Big] + \frac{1}{2} \muc \sqrt{\msd} \sin^2\phi^{}_1 \;. \hspace{0.8cm}
\label{eq:Ratio2}
\end{eqnarray}
One can see that $|V^{}_{td}/V^{}_{ts}| \simeq \sqrt{m^{}_d/m^{}_s}$ and
$|V^{}_{us}/V^{}_{ud}| \simeq \sqrt{m^{}_d/m^{}_s}$ hold at the leading order.
These two simple and instructive relations are well in accordance with current
experimental data. But a similar leading-order relation
$|V^{}_{ub}/V^{}_{cb}| \simeq \sqrt{m^{}_u/m^{}_c}$ is obviously in conflict with
the values of $m^{}_u$, $m^{}_c$, $|V^{}_{ub}|$ and $|V^{}_{cb}|$ given
in Eqs.~(\ref{eq:data-mass}) and (\ref{eq:data-mixing}), although it is
reasonably expected in the $m^{}_b \to \infty$ limit \cite{Xing:2012zv,Xing:2014sja}.
That is why the second term on the right-hand side of $|V^{}_{ub}/V^{}_{cb}|$
in Eq.~(\ref{eq:Ratio}) is crucial to fill in the gap and resolve the issue.

Let us proceed to make a ball-park estimate of the allowed ranges of those free
parameters appearing in Eq.~(\ref{eq:Ratio}).
If $|V^{}_{ub}/V^{}_{cb}| \simeq 0.089$, $\sqrt{m^{}_u/m^{}_c} \simeq 0.045$ and
$m^{}_s/m^{}_b \sqrt{m^{}_d/m^{}_s} \simeq 0.004$ are typically taken into account,
we find that
\begin{eqnarray}
\left| \sqrt{\rd{}} ~
\frac{\sqrt{\ru{}} \cos\phi^{}_2 - \sqrt{\rd{}} \cos\phi^{}_1}{\ru{}
+ \rd{} - 2\sqrt{\ru{}\rd{}} \cos\delta} \right| \sim 10
\label{eq:condition}
\end{eqnarray}
should hold so as to make the first formula in Eq.~(\ref{eq:Ratio}) valid. On the
other hand, we keep in mind that the relation
$|V^{}_{us}| \simeq \sqrt{m^{}_d/m^{}_s}$ coincides with the experimental
data very well. Thus the term proportional to $\sqrt{m^{}_u/m^{}_c}$ in the
expression of $|V^{}_{us}|$ as shown in Eq.~(\ref{eq:abs-Vckm}) must be very small.
This observation implies that $\cos\phi^{}_1 \sim 0$ is naturally expected.
Substituting $\cos\phi^{}_1 \sim 0$ into Eq.~(\ref{eq:condition}), one has
\begin{eqnarray}
\left|  \frac{\xi \cos\phi^{}_2}{1+\xi^2 \pm 2\xi \sin\phi^{}_2} \right| \sim 10 \;,
\label{eq:condition-app}
\end{eqnarray}
where $\xi \equiv \sqrt{\ru{}/\rd{}}$ is defined, and ``$\pm$" corresponds to
$\phi^{}_1 \sim \mp \pi/2$. The left-hand side of Eq.~(\ref{eq:condition-app})
is a function of $\phi^{}_2$ and takes its maximum value
$|\xi/\left(1-\xi^2\right)|$ at $\sin \phi^{}_2 = \pm 2\xi/\left(1+\xi^2\right)$.
As a consequence, the condition in Eq.~(\ref{eq:condition-app}) leads us to the
constraint $|\xi/\left(1-\xi^2\right)| \gtrsim 10$, or equivalently
$\left(\sqrt{401} -1\right)/20 \lesssim \xi \lesssim \left(\sqrt{401} + 1\right)/20$.
This result means that the ratio $\ru{}/\rd{}$ lies in the range of $0.9$ to $1.1$.
In other words, $\ru{} \simeq \rd{}$ is expected to hold as a good approximation.
This interesting point has been seen from a careful numerical analysis of the
parameter space of $\ru{}$ and $\rd{}$ in Ref.~\cite{Xing:2015sva}, but here we
reach the same observation based mainly on our reliable analytical approximations
made above.

\subsection{A special case with $r^{}_{\rm u} = r^{}_{\rm d}$}
\label{section:2.3}

Now that $r^{}_{\rm u} \simeq r^{}_{\rm d}$ is favored by current experimental
data (especially by today's experimental result for $|V^{}_{ub}/V^{}_{cb}|$), we are
well motivated to consider the special case $\ru{} = \rd{} \equiv r$ as another
natural consequence of the structural parallelism between the up- and down-quark
sectors. In this case Eq.~(\ref{eq:condition}) is simplified to
$\left|\sin\phi/\sin\left(\delta/2\right)\right| \sim 20$,
where $\phi \equiv \left(\phi^{}_1 + \phi^{}_2\right)/2$ has been defined.
Then we are left with $\phi^{}_2 \sim \phi^{}_1 \sim \pm \pi/2$, an estimate
consistent with Eq.~(\ref{eq:condition-app}) for $\xi = 1$. Moreover,
the expression of $R$ is simplified to $R = 4 r \sin^2\left(
\delta/2 \right)$, which can easily reach the level of $\mathcal{O}
\left(\lambda^4\right)$ for $r\sim \mathcal{O} \left( \lambda \right)$ due to the
smallness of $\delta = \phi^{}_1 - \phi^{}_2$ in magnitude. As a result, the
approximate formulas for the CKM matrix elements in Eq.~(\ref{eq:abs-Vckm})
are now simplified to
\begin{eqnarray}
|V^{}_{ud}| &=& 1 - \frac{1}{2} \mds + \eu\ed \sqrt{\muc} \sqrt{\mds} \cos\phi^{}_1 \;,
\nonumber
\\
|V^{}_{us}| &=& \sqrt{\mds} \left( 1 - \frac{1}{2} \mds \right) + \eu\ed \sqrt{\muc}
\Big[ \left( r - 1 \right) \cos\phi^{}_1 - r\cos\phi^{}_2 \Big] + \frac{1}{2}
\muc \sqrt{\msd} \sin^2\phi^{}_1 \;,
\nonumber
\\
|V^{}_{ub}| &=& \sqrt{r} \left[ \left( 2 - r \right) \sqrt{\muc} + \eu \msb \sqrt{\mds}
\frac{\sin\phi}{\displaystyle\sin\frac{\delta}{2}} \right]
\left|\sin\frac{\delta}{2}\right| \;,
\nonumber
\\
|V^{}_{cd}| &=& \sqrt{\mds} \left( 1 - 2r\sin^2 \frac{\delta}{2} + \frac{1}{2} r^2
\sin^2\delta - \frac{1}{2} \mds \right) + \frac{1}{2} \muc \sqrt{\msd} \sin^2\phi^{}_1
\nonumber
\\
&&  - \eu\ed \sqrt{\muc} \big( r \sin\phi^{}_1 \sin\delta + \cos\phi^{}_1 \big) \;,
\nonumber
\\
|V^{}_{cs}| &=& 1 - \frac{1}{2} \mds - r \left( 2 -\mds + \ed \msb
\right) \sin^2\frac{\delta}{2} + \frac{1}{2} r^2 \left( 1 + 2r \sin^2 \frac{\delta}{2}
\right) \sin^2\delta
\nonumber
\\
&& + \eu\ed \sqrt{\muc}\sqrt{\mds} \cos\phi^{}_1 \;,
\nonumber
\\
|V^{}_{cb}| &=& \sqrt{r} \left( 2 - r - \frac{1}{4} r^2 + \frac{1}{2} \ed \msb \right)
\left| \sin\frac{\delta}{2} \right| \;,
\nonumber
\\
|V^{}_{td}| &=& \sqrt{r} \sqrt{\mds} \left( 2 - r - \frac{1}{4} r^2 - \frac{1}{2}
\ed \msb - \mds \right) \left|\sin\frac{\delta}{2} \right| \;,
\nonumber
\\
|V^{}_{ts}| &=& \sqrt{r} \left( 2 - r - \frac{1}{4} r^2 + \frac{1}{2} \ed \msb
- \mds \right) \left| \sin\frac{\delta}{2} \right| \;,
\nonumber
\\
|V^{}_{tb}| &=& 1 - r \left( 2 + \ed \msb \right) \sin^2\frac{\delta}{2} + \frac{1}{2}
r^2 \left( 1 + 2r\sin^2\frac{\delta}{2} \right) \sin^2\delta \;.
\label{eq:abs-Vckm-app}
\end{eqnarray}
Accordingly, Eqs.~(\ref{eq:Ratio}) and (\ref{eq:Ratio2}) are simplified to
\begin{eqnarray}
\frac{|V^{}_{ub}|}{| V^{}_{cb} |} &=& \sqrt{\muc} + \frac{1}{2} \eu
\msb \sqrt{\mds} \frac{\sin\phi}{\displaystyle\sin\frac{\delta}{2}} \;, \hspace{0.3cm}
\nonumber
\\
\frac{|V^{}_{td}|}{| V^{}_{ts} |} &=& \sqrt{\mds} \left( 1 - \frac{1}{2} \ed \msb \right) \;,
\label{eq:Ratio-app}
\end{eqnarray}
and
\begin{eqnarray}
\frac{|V^{}_{us}|}{| V^{}_{ud} |} &=& \sqrt{\mds} + \eu \ed \sqrt{\muc}
\Big[ \left(r - 1 \right) \cos\phi^{}_1 - r\cos\phi^{}_2 \Big]
+ \frac{1}{2} \muc \sqrt{\msd} \sin^2\phi^{}_1 \;,
\label{eq:Ratio-app2}
\end{eqnarray}
respectively. Some brief comments are in order.
\begin{itemize}
\item      One can see that the expressions of $|V^{}_{ub}|$, $|V^{}_{cb}|$, $|V^{}_{td}|$
and $|V^{}_{ts}|$ are all approximately proportional to $|\sin\left( \delta/2 \right)|$
in this special case. This result is exactly a consequence of the up-down structural
parallelism, implying some significant cancellations between the two sectors so as to
arrive at sufficiently small values for the four off-diagonal CKM matrix elements
associated with the third-family quarks. Note, however, that
$\sin\left(\delta/2\right)$ should not be too small, in order to assure
$|V^{}_{cb}|$ and $|V^{}_{ts}|$ to reach the level of ${\cal O}\left(\lambda^2\right)$.

\item     Eq.~(\ref{eq:abs-Vckm-app}) shows that $|V^{}_{tb}| > |V^{}_{cs}|$,
$|V^{}_{ud}| > |V^{}_{cs}|$ and $|V^{}_{cb}| > |V^{}_{ts}|$ hold as a natural
result of the smallness of $r$ and $\delta$. These fine inequalities reveal some
details of the obtained pattern of the unitary CKM matrix $V$, and they are fully
consistent with the experimental data listed in Eq.~(\ref{eq:data-mixing})
\cite{Zyla:2020zbs}.

\item     It is worth pointing out that the analytical approximations
given in Eqs.~(\ref{eq:abs-Vckm-app})---(\ref{eq:Ratio-app2}) keep unchanged
under the replacements $\eu \to - \eu$, $\ed \to \ed$ and
$\phi^{}_{1,2} \to \phi^{}_{1,2} \pm \pi$. This observation
means that once $\ed$ is fixed (i.e., either $\ed =+1$ or $-1$), the $\eu = \pm 1$
cases will lead to the same results for the moduli of nine CKM matrix elements with
the help of Eqs.~(\ref{eq:abs-Vckm-app})---(\ref{eq:Ratio-app2}). The two independent
phases in these two cases differ from each other by $\pi$.
One may easily check that such a conclusion is also valid for the generic
analytical approximations made in the $ \ru{} \neq \rd{}$ case, as shown
in Eq.~(\ref{eq:abs-Vckm}). 
But one should keep in mind that the validity of the above 
observations is essentially a consequence of $\left(O^{}_{\rm u} \right)^{}_{1t} \simeq 0$ 
up to $\mathcal{O} \left(\lambda^4 \right)$ in Eq.~(\ref{eq:app-u}),
which leads us to $|V^{}_{t q}| \simeq \big|\left(O^{}_{\rm u}\right)^{}_{2 t}
\left(O^{}_{\rm d}\right)^{}_{2 q} e^{{\rm i}\phi^{}_1} + \left(O^{}_{\rm u}\right)^{}_{3 t}
\left(O^{}_{\rm d}\right)^{}_{3 q} e^{{\rm i}\phi^{}_2}\big|$ (for $q=d,s,b$) as can be 
seen from Eq.~(\ref{eq:Vckm}).
In view of the structural parallelism between $M^{}_{\rm u}$ and $M^{}_{\rm d}$,
one might naively expect that a similar conclusion could be drawn for the $\ed = \pm 1$
cases against a fixed choice of $\eu$. However, it is indeed not the case because
the much weaker hierarchy of three down-type quark masses results in
$\left( O^{}_{\rm d} \right)^{}_{1b} \sim \mathcal{O} \left( \lambda^3 \right)$ as 
shown in Eq.~(\ref{eq:app-d}). Hence $|V^{}_{pb}|$ is sensitive to the contribution 
from $\left( O^{}_{\rm u} \right)^{}_{1p} \left( O^{}_{\rm d} \right)^{}_{1b}$
(for $p=u,c,t$) to some extent in our analytical approximations.
\end{itemize}
In short, the {\it Ansatz} based on the assumption of $r^{}_{\rm u} = r^{}_{\rm d}$
might be interesting from the point of view of model building with the help of a
kind of flavor symmetry, but an apparent asymmetry between the mass hierarchies
of up- and down-type quarks implies that $r^{}_{\rm u} \simeq r^{}_{\rm d}$ should
be more reasonable and thus more likely. Of course, why $m^{}_t/m^{}_b \sim 60$,
$m^{}_c/m^{}_s \sim 12$ but $m^{}_u/m^{}_d \sim 0.5$ hold at a given energy scale
remains a big puzzle in particle physics, especially in view of the fact that
the masses of up- and down-type quarks originate from the similar
Yukawa interactions in the SM and its natural extensions.

\subsection{Textures of $M^{}_{\rm u,d}$ and CP violation}

Once the free parameters $r^{}_{\rm u}$ and $r^{}_{\rm d}$ are well constrained by
current experimental data on the CKM matrix elements and quark masses based on
the zero textures of $M^{}_{\rm u}$ and $M^{}_{\rm d}$,
one will be able to determine or constrain
those nonzero elements of $M^{}_{\rm u}$ and $M^{}_{\rm d}$ to gain a deeper
understanding of their hierarchical textures. So let us
go back to Eq.~(\ref{eq:mass-para}) and reexpress those real matrix elements as
follows:
\begin{eqnarray}
&& \frac{|C^{}_{\rm u}|}{m^{}_t} =  \mct \sqrt{\frac{1}{1-\ru{}}\muc} \;,
\nonumber
\\
&& \frac{\tilde{B}^{}_{\rm u}}{m^{}_t} = \ru{} - \eu \frac{m^{}_u}{m^{}_t} + \eu \mct \;,
\nonumber
\\
&& \frac{|B^{}_{\rm u}|}{m^{}_t} = \sqrt{\ru{} \left( 1 - \ru{} + \eu \frac{m^{}_u}{m^{}_t}
- \eu \mct - \frac{1}{1-\ru{}} \frac{m^{}_u m^{}_c}{m^2_t} \right) } \;, \hspace{0.5cm}
\label{eq:mpu}
\end{eqnarray}
and
\begin{eqnarray}
&& \frac{|C^{}_{\rm d}|}{m^{}_b} = \msb \sqrt{\frac{1}{1-\rd{}}\mds} \;,
\nonumber
\\
&& \frac{\tilde{B}^{}_{\rm d}}{m^{}_b} = \rd{} - \ed \frac{m^{}_d}{m^{}_b} + \ed \msb \;,
\nonumber
\\
&& \frac{|B^{}_{\rm d}|}{m^{}_b} = \sqrt{\rd{} \left( 1 - \rd{} + \ed \frac{m^{}_d}{m^{}_b}
- \ed \msb - \frac{1}{1-\rd{}} \frac{m^{}_d m^{}_s}{m^2_b} \right) } \;. \hspace{0.5cm}
\label{eq:mpd}
\end{eqnarray}
It is then straightforward to obtain the approximate analytical results of these
parameters by making appropriate expansions of the right-hand sides of
Eqs.~(\ref{eq:mpu}) and (\ref{eq:mpd}) in terms of the quark mass ratios and
$r^{}_{\rm q}$ (for $\rm q = u, d$). Here we focus on the ratios
$\tilde{B}^{}_{\rm u}/|B^{}_{\rm u}|$ and $\tilde{B}^{}_{\rm d}/|B^{}_{\rm d}|$,
in order to understand why fitting the experimental result of $|V^{}_{ub}/V^{}_{cb}|$
requires the zero textures of $M^{}_{\rm u}$ and $M^{}_{\rm d}$ to have
a somewhat weak hierarchy on their right-bottom corners.

To be explicit, we have
\begin{eqnarray}
&& \frac{\tilde{B}^{}_{\rm u}}{|B^{}_{\rm u}|} \simeq \sqrt{\ru{}} \left( 1 + \frac{1}{2}
\ru{} + \frac{3}{8} \ru{2} \right) \;,
\nonumber
\\
&& \frac{\tilde{B}^{}_{\rm d}}{|B^{}_{\rm d}|} \simeq \sqrt{\rd{}} \left[ 1 +
\left(\frac{1}{2} \rd{} + \frac{\ed}{\rd{}} \msb \right) + \left( \frac{3}{8}\rd{2}
+ \ed \msb \right) \right]  \;, \hspace{0.6cm}
\label{eq:RB-mass}
\end{eqnarray}
up to the accuracy of $\mathcal{O} \left( \lambda^3 \right)$ for
$r^{}_{\rm u} \sim r^{}_{\rm d} \sim {\cal O}\left(\lambda\right)$. Given the
fact that $r^{}_{\rm u}$ and $r^{}_{\rm d}$ measure the deviations of $A^{}_{\rm u}$
and $A^{}_{\rm d}$ respectively from $m^{}_t$ and $m^{}_b$, we arrive at
\begin{eqnarray}
\frac{\tilde{B}^{}_{\rm u}}{|B^{}_{\rm u}|} \sim \frac{|B^{}_{\rm u}|}{A^{}_{\rm u}}
\sim \sqrt{\ru{}} \; , \quad
\frac{\tilde{B}^{}_{\rm d}}{|B^{}_{\rm d}|} \sim \frac{|B^{}_{\rm d}|}{A^{}_{\rm d}}
\sim \sqrt{\rd{}} \; . \hspace{0.3cm}
\label{eq:23texture}
\end{eqnarray}
This approximate observation implies that the four nonzero elements on the right-bottom
corner of $M^{}_{\rm q}$ (for $\rm q = u$ or $\rm d$) should not have a very strong
hierarchy in magnitude. Instead, their values satisfy an approximate ``seesaw"
relation $\tilde{B}^{}_{\rm q} \sim |B^{}_{\rm q}|^2/A^{}_{\rm q}$. Although such a
relation has been observed before \cite{Fritzsch:2002ga,Xing:2015sva}, our present
analytical approximations provide a much better understanding of why this is the case.
If the {\it Ansatz} with $\ru{} = \rd{} \equiv r$ is taken into account,
one will be similarly left with $\tilde{B}^{}_{\rm u}/|B^{}_{\rm u}| \sim
\tilde{B}^{}_{\rm d}/|B^{}_{\rm d}| \sim \sqrt{r}$.

Let us highlight that the seesaw-like relation
$\tilde{B}^{}_{\rm q} \sim |B^{}_{\rm q}|^2/A^{}_{\rm q}$ associated with the
structure of $M^{}_{\rm q}$ is actually a phenomenological compromise between
resolving the failure of the popular relation
$|V^{}_{ub}/V^{}_{cb}| \simeq \sqrt{m^{}_u/m^{}_c}$ and
generating a sufficiently strong quark mass hierarchy (i.e., $m^{}_c \ll m^{}_t$ and
$m^{}_s \ll m^{}_b$). In this way we are left with
a relatively weak texture hierarchies on the right-bottom corners of
$M^{}_{\rm u}$ and $M^{}_{\rm d}$, and achieve a sufficiently large correction
to $|V^{}_{ub}/V^{}_{cb}|$ from the down-type quark sector as shown in
Eq.~(\ref{eq:Ratio}) or Eq.~(\ref{eq:Ratio-app}).

It is well known that the strength of CP violation in the quark sector can be
described by the rephasing-invariant Jarlskog parameter ${\cal J}$~\cite{Jarlskog:1985ht},
which is defined by
\begin{eqnarray}
{\cal J} \sum_\gamma \varepsilon^{}_{\alpha\beta\gamma} \sum^{}_k
\varepsilon^{}_{ijk} = {\rm Im} \left(V^{}_{\alpha i} V^{}_{\beta j}
V^*_{\alpha j} V^*_{\beta i}\right) \; ,
\label{eq:J}
\end{eqnarray}
where the Greek and Latin subscripts run over the up- and down-type quark flavors,
respectively. Taking advantage of the analytical approximations made for the
CKM matrix elements in Eq. (\ref{eq:ele-Vckm}),
we obtain the approximate expression of the Jarlskog invariant as follows:
\begin{eqnarray}
\mathcal{J} &\simeq& \eu \sqrt{\ru{}\rd{}} \hspace{0.1cm}
\frac{m^{}_d}{m^{}_b} \sin\delta
- \frac{1}{2} \eu \ed \sqrt{\muc} \sqrt{\mds} \hspace{0.1cm}
\Big\{ \Big[ \left( \ru{} + \rd{} - 2 \right) \left( \ru{} + \rd{}
- 4\sqrt{\ru{}\rd{}} \cos\delta \right) \hspace{0.3cm}
\nonumber
\\
&& + 4\sqrt{\ru{}\rd{}} \left( \sqrt{\ru{}\rd{}} - \cos\delta \right) \Big]
\sin\phi^{}_1 - 2 \sqrt{\ru{}\rd{}} \left( \ru{} + \rd{} - 2\sqrt{\ru{}\rd{}}
\cos\delta\right) \sin\phi^{}_2 \Big\} \; ,
\label{eq:Jarlskog}
\end{eqnarray}
in which $\delta$ is correlated with $\phi^{}_1$ and $\phi^{}_2$ through
$\delta = \phi^{}_1 - \phi^{}_2$. If the {\it Ansatz} with
$r^{}_{\rm u} = r^{}_{\rm d}$ is taken into consideration,
then Eq.~(\ref{eq:Jarlskog}) can be simplified to
\begin{eqnarray}
\mathcal{J}\left(r^{}_{\rm u} = r^{}_{\rm d} \equiv r\right)
\simeq 4 \eu\ed r \sqrt{\muc} \sqrt{\mds} \Big[ \left( 1 - 2r \right)
\sin\phi^{}_1 + r\sin\phi^{}_2 \Big] \sin^2\frac{\delta}{2} + \ed r
\frac{m^{}_d}{m^{}_b} \sin\delta \;.
\label{eq:Jarlskog-app}
\end{eqnarray}
The two terms on the right-hand side of Eq.~(\ref{eq:Jarlskog-app}) should
contribute comparably to the Jarlskog invariant, because both $\delta$ and $r$
are small. In this case the observed positivity of $\mathcal{J}$ \cite{Zyla:2020zbs}
requires $\eu\ed \sin\phi^{}_1 > 0$ and $\ed \sin\delta >0$. To make the result of
$|V^{}_{ub}|/|V^{}_{cb}|$ in Eq.~(\ref{eq:Ratio-app}) coincide with the experimental
data, $\eu \sin \phi / \sin\left(\delta/2\right) > 0$ is demanded.
Similarly, the result of $|V^{}_{us}|$ given in
Eq.~(\ref{eq:abs-Vckm-app}) requires $\eu\ed\cos\phi^{}_1 < 0$. For
$\phi^{}_1 \sim \phi^{}_2 \sim \pm \pi/2$, we find that $\sin\phi^{}_1$,
$\sin\phi^{}_2$ and $\sin\phi$ should all have the same sign. Thus we have
\begin{eqnarray}
&&\phi^{}_1 \sim \phi^{}_2 \sim \frac{\pi}{2}~{\rm or}~ \frac{3\pi}{2}  \;,\quad
\phi^{}_1 \gtrsim \frac{\pi}{2}~{\rm or}~ \frac{3\pi}{2} \;,\quad
\eu\ed \sin \phi^{}_1 >0 \;,\quad \ed \delta > 0 \;,
\label{eq:phases-relation}
\end{eqnarray}
and $\eu \sin\phi/ \sin \left(\delta/2\right) > 0$ is accordingly
satisfied. The constraints obtained in Eq.~(\ref{eq:phases-relation})
are fully consistent with those obtained in Ref.~\cite{Xing:2015sva}, where
different phase conventions were used.

\section{Numerical results}

With the help of the approximate expressions of $|V^{}_{us}|$, $|V^{}_{ub}|$ and $|V^{}_{cb}|$
given in Eq.~(\ref{eq:abs-Vckm-app}) and the experimental data shown in Eq.~(\ref{eq:data-mixing}),
one may easily determine the values of $r$, $\phi^{}_1$ and $\phi^{}_2$ for the {\it Ansatz}
considered in section \ref{section:2.3}. In this regard
the positivity of ${\cal J}$ in Eq.~(\ref{eq:Jarlskog-app}) can be used to resolve possible
ambiguities associated with the two phase parameters. For the sake of simplicity, here we only
adopt the central values of $|V^{}_{us}|$, $|V^{}_{ub}|$, $|V^{}_{cb}|$
and six quark masses listed in Eqs.~(\ref{eq:data-mass}) and (\ref{eq:data-mixing}) in
our numerical calculations.

\subsection{$\eu =+1$ and $\ed = + 1$}

In this case we have $(r, \phi^{}_1, \phi^{}_2) = (0.18556, 0.53216 \pi, 0.49914 \pi )$.
Then we arrive at
        \begin{eqnarray}
        |V| \simeq \begin{pmatrix}
        0.97388 & \;\;\; 0.22650 \;\;\; & 0.00361
        \\
        0.22632 & \;\;\; 0.97308 \;\;\; & 0.04053
        \\
        0.00874 & \;\;\; 0.03941 \;\;\; & 0.99918
        \end{pmatrix}\;,
        \label{eq:V-case1}
        \end{eqnarray}
        and
        \begin{eqnarray}
        \frac{|V^{}_{ub}|}{|V^{}_{cb}|} \simeq 0.08497 \;,\quad \frac{|V^{}_{td}|}{|V^{}_{ts}|}
        \simeq  0.22201 \;,\quad \frac{|V^{}_{us}|}{|V^{}_{ud}|} \simeq 0.23213 \;,
        \label{eq:R-case1}
        \end{eqnarray}
        where Eqs.~(\ref{eq:abs-Vckm-app}), (\ref{eq:Ratio-app}) and (\ref{eq:Ratio-app2})
        have been used. If the above three ratios are directly calculated from the results
        in Eq.~(\ref{eq:V-case1}), their values
        will deviate from those obtained in Eq.~(\ref{eq:R-case1})
        by $4.6\%$, $0.13\%$ and $0.19\%$, respectively. Using Eq.~(\ref{eq:Jarlskog-app})
        to calculate the Jarlskog invariant, we obtain a reasonable value
        $\mathcal{J} \simeq 3.42 \times 10^{-5}$.

To check the accuracy of our analytical approximations, we recalculate the moduli of
nine CKM matrix elements by means of Eqs.~(\ref{eq:orth}) and (\ref{eq:Vckm}) in the
case of $r^{}_{\rm u} = r^{}_{\rm d} \equiv r$. We find that the numerical results
obtained in this way (named as ``exact") deviate slightly from those achieved from the
analytical approximations in Eq.~(\ref{eq:V-case1}):
        \begin{eqnarray}
        \frac{|V|^{}_{\rm approx}- |V|^{}_{\rm exact}}{|V|^{}_{\rm exact}}
        \simeq \begin{pmatrix}
        -0.03149 & \;\;\; +0.35053 \;\;\; & -5.70473
        \\
        +0.33591 & \;\;\; -0.03112 \;\;\; & +0.11371
        \\
        -2.12646 & \;\;\; -0.66504 \;\;\; & +0.00054
        \end{pmatrix} \% \;.
        \label{eq:error-case1}
        \end{eqnarray}
It is obvious that the two smallest CKM matrix elements (namely, $|V^{}_{ub}|$ and
$|V^{}_{td}|$ involve the largest uncertainties, but the relative errors are at most
at the level of ${\cal O}\left(6\%\right)$. Moreover, the corresponding value of the
Jarlskog invariant is found to be $\mathcal{J}^{}_{\rm exact} \simeq 3.29 \times 10^{-5}$,
and the corresponding textures of quark mass matrices read as follows:
        \begin{eqnarray}
        \overline{M}^{}_{\rm u} &\simeq& m^{}_t \begin{pmatrix}
        0 & \;\;\; 0.00018 \;\;\; & 0
        \\
        0.00018 & \;\;\; 0.18924 \;\;\; & 0.38787
        \\
        0 & \;\;\; 0.38787 \;\;\; & 0.81444
        \end{pmatrix} \;,
        \nonumber
        \\
        \overline{M}^{}_{\rm d} &\simeq& m^{}_b \begin{pmatrix}
        0 & \;\;\; 0.00465 \;\;\; & 0
        \\
        0.00465 & \;\;\; 0.20335 \;\;\; & 0.38448
        \\
        0 & \;\;\; 0.38448 \;\;\; & 0.81444
        \end{pmatrix} \;,
        \label{eq:mm-case1}
        \end{eqnarray}
        which lead us to $\tilde{B}^{}_{\rm u}/|B^{}_{\rm u}| \simeq 0.488$ and $\tilde{B}^{}_{\rm d}/|B^{}_{\rm d}|
        \simeq 0.529$. The values of these two ratios are quite close to $1/2$,
        an interesting possibility which has been discussed by one of us in
        Ref.~\cite{Fritzsch:2021lzb}.

        Eq.~(\ref{eq:mm-case1}) clearly shows that the texture hierarchy associated
        with the right-bottom corner of $M^{}_{\rm q}$ is somewhat weaker than
        naively expected, and an approximate seesaw-like relation
        $\tilde{B}^{}_{\rm q} \sim |B^{}_{\rm q}|^2/A^{}_{\rm q}$ holds
        (for $\rm q = u$ or $\rm d$). One can therefore understand why those
        previous studies in the seemingly ``natural" assumption of $A^{}_{\rm q} \gg
        |B^{}_{\rm q}| \gg \tilde{B}^{}_{\rm q}$ (see, e.g.,
        Refs.~\cite{Du:1992iy,Fritzsch:1995nx,Xing:1996hi,Fritzsch:2021lzb,Kang:1997uv,
        Branco:1999nb,Matsuda:2006xa,Hernandez-Sanchez:2011xaa,Gupta:2013yha,
        Felix-Beltran:2013tra,Yang:2020qsa,Yang:2020goc,Yang:2021xob}) are invalid
        today to explain the experimental result for $|V^{}_{ub}/V^{}_{cb}|$.

\subsection{$\eu = + 1$ and $\ed = - 1$}

        In this case we have $(r, \phi^{}_1, \phi^{}_2) = (0.18371, 1.51969 \pi, 1.55317 \pi )$.
        Then we obtain
        \begin{eqnarray}
        |V| \simeq \begin{pmatrix}
        0.97427 & \;\;\; 0.22650 \;\;\; & 0.00361
        \\
        0.22632 & \;\;\; 0.97348 \;\;\; & 0.04053
        \\
        0.00892 & \;\;\; 0.03940 \;\;\; & 0.99918
        \end{pmatrix}\;,
        \label{eq:V-case2}
        \end{eqnarray}
        and
        \begin{eqnarray}
        \frac{|V^{}_{ub}|}{|V^{}_{cb}|} \simeq 0.08419 \;,\quad \frac{|V^{}_{td}|}{|V^{}_{ts}|}
        \simeq 0.22621 \;,\quad \frac{|V^{}_{us}|}{|V^{}_{ud}|} \simeq 0.23213 \;,
        \label{eq:R-case2}
        \end{eqnarray}
        where Eqs.~(\ref{eq:abs-Vckm-app}), (\ref{eq:Ratio-app}) and (\ref{eq:Ratio-app2})
        have been used. If the above three ratios are simply calculated from the results in Eq.~(\ref{eq:V-case2}), their values will deviate from those give in Eq.~(\ref{eq:R-case2}) by $5.5\%$, $0.13\%$ and $0.15\%$, respectively. As for
        the Jarlskog invariant, we arrive at $\mathcal{J} \simeq 3.46 \times 10^{-5}$
        by means of the approximate formula given in Eq.~(\ref{eq:Jarlskog-app}).

        Taking $r^{}_{\rm u} = r^{}_{\rm d} \equiv r$ and using Eqs.~(\ref{eq:orth}) and (\ref{eq:Vckm}), we recalculate the moduli of nine CKM matrix elements and
        compare the results with those approximate ones obtained in Eq.~(\ref{eq:V-case2}):
        \begin{eqnarray}
        \frac{|V|^{}_{\rm approx}- |V|^{}_{\rm exact}}{|V|^{}_{\rm exact}}
        \simeq \begin{pmatrix}
        +0.02942 & \;\;\; -0.03841 \;\;\; & -4.50882
        \\
        -0.05558 & \;\;\; +0.03143 \;\;\; & -0.64582
        \\
        -0.01523 & \;\;\; -1.46565 \;\;\; & +0.00196
        \end{pmatrix} \% \;.
        \label{eq:error-case2}
        \end{eqnarray}
        In addition, we obtain $\mathcal{J}^{}_{\rm exact} \simeq 3.26 \times 10^{-5}$
        for the Jarlskog invariant. The textures of quark mass matrices are found to be
        \begin{eqnarray}
        \overline{M}^{}_{\rm u} &\simeq& m^{}_t \begin{pmatrix}
        0 & \;\;\; 0.00018 \;\;\; & 0
        \\
        0.00018 & \;\;\; 0.18739 \;\;\; & 0.38638
        \\
        0 & \;\;\; 0.38638 \;\;\; & 0.81629
        \end{pmatrix} \;,
        \nonumber
        \\
        \overline{M}^{}_{\rm d} &\simeq& m^{}_b \begin{pmatrix}
        0 & \;\;\; 0.00464 \;\;\; & 0
        \\
        0.00464 & \;\;\; 0.16593 \;\;\; & 0.39144
        \\
        0 & \;\;\; 0.39144 \;\;\; & 0.81629
        \end{pmatrix} \;,
        \label{eq:mm-case2}
        \end{eqnarray}
        from which we get $\tilde{B}^{}_{\rm u}/|B^{}_{\rm u}| \simeq 0.485$ and $\tilde{B}^{}_{\rm d}/|B^{}_{\rm d}| \simeq 0.424$.

\subsection{$\eu = - 1$ and $\ed = + 1$}

        In this case we have $(r, \phi^{}_1, \phi^{}_2) = (0.18556, 1.53216 \pi, 1.49914 \pi )$.
        Then we are left with
        \begin{eqnarray}
        |V| \simeq \begin{pmatrix}
        0.97388 & \;\;\; 0.22650 \;\;\; & 0.00361
        \\
        0.22632 & \;\;\; 0.97308 \;\;\; & 0.04053
        \\
        0.00874 & \;\;\; 0.03941 \;\;\; & 0.99918
        \end{pmatrix}\;,
        \label{eq:V-case3}
        \end{eqnarray}
        and
        \begin{eqnarray}
        \frac{|V^{}_{ub}|}{|V^{}_{cb}|} \simeq 0.08497 \;, \quad \frac{|V^{}_{td}|}{|V^{}_{ts}|}
        \simeq 0.22201 \;, \quad \frac{|V^{}_{us}|}{|V^{}_{ud}|} \simeq 0.23213 \;,
        \label{eq:R-case3}
        \end{eqnarray}
        where Eqs.~(\ref{eq:abs-Vckm-app}), (\ref{eq:Ratio-app}) and (\ref{eq:Ratio-app2})
        have been used. If the above three ratios are directly calculated by using
        the results obtained in Eq.~(\ref{eq:V-case3}), their values will deviate
        from those given in Eq.~(\ref{eq:R-case3}) by $4.6\%$, $0.13\%$ and $0.19\%$,
        respectively. Moreover, we arrive at $\mathcal{J} \simeq 3.42 \times 10^{-5}$.

        Taking $r^{}_{\rm u} = r^{}_{\rm d} \equiv r$ and using Eqs.~(\ref{eq:orth}) and (\ref{eq:Vckm}), we recalculate the moduli of nine CKM matrix elements and
        compare the results with those approximate ones obtained in
        Eq.~(\ref{eq:V-case3}):
        \begin{eqnarray}
        \frac{|V|^{}_{\rm approx}- |V|^{}_{\rm exact}}{|V|^{}_{\rm exact}} \simeq
        \begin{pmatrix}
        -0.03134 & \;\;\; +0.34760 \;\;\; & -5.65788
        \\
        +0.32934 & \;\;\; -0.03021 \;\;\; & -0.20508
        \\
        +0.22741 & \;\;\; -1.10930 \;\;\; & +0.00107
        \end{pmatrix} \% \;.
        \label{eq:error-case3}
        \end{eqnarray}
        In addition, we obtain
        $\mathcal{J}^{}_{\rm exact} \simeq 3.23 \times 10^{-5}$ and
        \begin{eqnarray}
        \overline{M}^{}_{\rm u} &\simeq& m^{}_t \begin{pmatrix}
        0 & \;\;\; 0.00018 \;\;\; & 0
        \\
        0.00018 & \;\;\; 0.18189 \;\;\; & 0.38963
        \\
        0 & \;\;\; 0.38963 \;\;\; & 0.81444
        \end{pmatrix} \;,
        \nonumber
        \\
        \overline{M}^{}_{\rm d} &\simeq& m^{}_b \begin{pmatrix}
        0 & \;\;\; 0.00465 \;\;\; & 0
        \\
        0.00465 & \;\;\; 0.20335 \;\;\; & 0.38448
        \\
        0 & \;\;\; 0.38448 \;\;\; & 0.81444
        \end{pmatrix} \;,
        \label{eq:mm-case3}
        \end{eqnarray}
        from which $\tilde{B}^{}_{\rm u}/|B^{}_{\rm u}| \simeq 0.467$ and
        $\tilde{B}^{}_{\rm d}/|B^{}_{\rm d}| \simeq 0.529$ can be extracted.

\subsection{$\eu = - 1$ and $\ed = - 1$}

        In this case we have $(r, \phi^{}_1, \phi^{}_2) = (0.18371, 0.51969 \pi, 0.55318 \pi)$
        and obtain
        \begin{eqnarray}
        |V| \simeq \begin{pmatrix}
        0.97427 & \;\;\; 0.22650 \;\;\; & 0.00361
        \\
        0.22632 & \;\;\; 0.97348 \;\;\; & 0.04053
        \\
        0.00892 & \;\;\; 0.03940 \;\;\; & 0.99918
        \end{pmatrix}\;,
        \label{eq:V-case4}
        \end{eqnarray}
        and
        \begin{eqnarray}
        \frac{|V^{}_{ub}|}{|V^{}_{cb}|} \simeq 0.08419 \;,
        \quad \frac{|V^{}_{td}|}{|V^{}_{ts}|}
        \simeq 0.22621 \;, \quad \frac{|V^{}_{us}|}{|V^{}_{ud}|} \simeq 0.23213 \;,
        \label{eq:R-case4}
        \end{eqnarray}
        where Eqs.~(\ref{eq:abs-Vckm-app}), (\ref{eq:Ratio-app}) and (\ref{eq:Ratio-app2})
        have been used. If the above three ratios are simply calculated
        by means of the results given in Eq.~(\ref{eq:V-case4}), they will
        deviate from those obtained in Eq.~(\ref{eq:R-case4}) by $5.5\%$, $0.13\%$ and $0.15\%$,
        respectively. As for the Jarlskog invariant of CP violation, we get
        $\mathcal{J} \simeq 3.46 \times 10^{-5}$ with the help of
        Eq.~(\ref{eq:Jarlskog-app}).

        Taking $r^{}_{\rm u} = r^{}_{\rm d} \equiv r$ and using Eqs.~(\ref{eq:orth}) and (\ref{eq:Vckm}), we recalculate the moduli of nine CKM matrix elements and
        compare the results with those approximate ones obtained in
        Eq.~(\ref{eq:V-case4}):
        \begin{eqnarray}
        \frac{|V|^{}_{\rm approx}- |V|^{}_{\rm exact}}{|V|^{}_{\rm exact}}
        \simeq \begin{pmatrix}
        +0.02972 & \;\;\; -0.04378 \;\;\; & -4.82382
        \\
        -0.05785 & \;\;\; +0.03054 \;\;\; & -0.06763
        \\
        -2.01585 & \;\;\; -0.76827 \;\;\; & +0.00101
        \end{pmatrix} \% \;.
        \label{eq:error-case4}
        \end{eqnarray}
        Furthermore, we obtain $\mathcal{J}^{}_{\rm exact} \simeq 3.31 \times 10^{-5}$
        and
        \begin{eqnarray}
        \overline{M}^{}_{\rm u} &\simeq& m^{}_t \begin{pmatrix}
        0 & \;\;\; 0.00018 \;\;\; & 0
        \\
        0.00018 & \;\;\; 0.18003 \;\;\; & 0.38812
        \\
        0 & \;\;\; 0.38812 \;\;\; & 0.81629
        \end{pmatrix} \;,
        \nonumber
        \\
        \overline{M}^{}_{\rm d} &\simeq& m^{}_b \begin{pmatrix}
        0 & \;\;\; 0.00464 \;\;\; & 0
        \\
        0.00464 & \;\;\; 0.16593 \;\;\; & 0.39144
        \\
        0 & \;\;\; 0.39144 \;\;\; & 0.81629
        \end{pmatrix} \;,
        \label{eq:mm-case4}
        \end{eqnarray}
        from which $\tilde{B}^{}_{\rm u}/|B^{}_{\rm u}| \simeq 0.464$ and
        $\tilde{B}^{}_{\rm d}/|B^{}_{\rm d}| \simeq 0.424$ can be achieved.

Our numerical results confirm that the moduli of nine CKM matrix elements and
the Jarlskog invariant of CP violation obtained from our analytical approximations
are the same in the cases with $\ed=+1$ (or $\ed = -1$) and $\eu = \pm 1$,
as pointed out in section~\ref{section:2.3}. 
As mentioned in section~\ref{section:2.3}, however, this conclusion 
only approximately holds. By comparing between Eqs.~(\ref{eq:error-case1}) and 
(\ref{eq:error-case3}) or between Eqs.~(\ref{eq:error-case2})
and (\ref{eq:error-case4}), one can see that the moduli of nine CKM matrix elements
calculated exactly by means of Eqs.~(\ref{eq:orth}) and (\ref{eq:Vckm}) are slightly
different from each other in the cases of $\ed=+1$ (or $\ed = -1$) and $\eu = \pm 1$.
Moreover, the nonzero elements of up- and down-type quark mass matrices are
not exactly the same in the above four cases. That is why in the beginning of
section~\ref{section:2} we have pointed out that different choices of the 
values of $\eta^{}_{\rm u}$ and $\eta^{}_{\rm d}$ correspond to different flavor
bases and thus to slightly different models of quark mass matrices.

\section{Conclusions}

Since 1977, many attempts have been made towards understanding the zero textures
of quark mass matrices and their connections with the CKM matrix elements.
Imposing the Hermiticity and parallelism on the up- and down-type quark sectors,
we are now left with the zero textures of $M^{}_{\rm u}$ and $M^{}_{\rm d}$
shown in Eq.~(\ref{eq:mass-matrix}) as the simplest viable possibility. To fit
current experimental data on quark flavor mixing and CP violation (especially
the observed value of $|V^{}_{ub}/V^{}_{cb}|$), however, we have analytically 
demonstrated that 
the $(2,2)$, $(2,3)$ and $(3,3)$ elements
of $M^{}_{\rm q}$ should
have a relatively weak texture hierarchy, much weaker than the corresponding
quark mass hierarchy. That is why the three relevant nonzero elements of
$M^{}_{\rm q}$ satisfy an approximate seesaw-like relation
$\tilde{B}^{}_{\rm q} \sim |B^{}_{\rm q}|^2/A^{}_{\rm q}$ (for $\rm q = u$ or
$\rm d$), and $|V^{}_{ub}/V^{}_{cb}|$ receives a quite large correction from
the down-type quark sector.

We highlight that our reliable analytical approximations have helped quite a lot
in resolving the phenomenological failure of the popular relation
$|V^{}_{ub}/V^{}_{cb}| \simeq \sqrt{m^{}_u/m^{}_c}$ and achieving a better
understanding of the correlation between quark mass and flavor mixing
hierarchies. In particular, we have clarified a somewhat misleading point of
view that a strong quark {\it mass} hierarchy ``naturally" corresponds to a
comparably strong {\it texture} hierarchy of the quark mass matrix. We emphasize
that the same caution should be exercised when studying the flavor issues of
charged leptons and massive neutrinos.

\section*{Acknowledgements}

This work was supported in part by the National Natural
Science Foundation of China under grant No. 12075254, grant
No. 11775231 and grant No. 11835013.

\end{document}